\documentclass[aps,prd,superscriptaddress,showpacs,amsmath,amssymb,amsfonts,nofootinbib,showpacs,reprint]{revtex4-1}

\usepackage{graphicx}
\usepackage{color}
\usepackage{slashed}
\usepackage{tabularx}
\usepackage[colorlinks = true,linkcolor=blue,citecolor = blue]{hyperref}
\usepackage{bbold}
\usepackage{bbm}
\usepackage{amsmath}
\usepackage{placeins}
\usepackage[normalem]{ulem}
\usepackage{dsfont}
\usepackage{cleveref}
\usepackage{comment}

\newcommand{\iu}{\ensuremath{\mathrm{i}}}

\newcommand{\idp}{\int\!\frac{d^3p}{(2\pi)^3\!}\,}

\newcommand{\muq}{\mu_q}
\newcommand{\muI}{\mu_I}
\newcommand{\muRG}{\nu}

\begin{document}

\newcommand{\BU}{Fakult\"at f\"ur Physik, Universit\"at Bielefeld, 33615 Bielefeld, Germany}
\newcommand{\GU}{Institut f\"{u}r Theoretische Physik, Goethe-Universit\"{a}t Frankfurt,
 Max-von-Laue-Str.\ 1, 60438 Frankfurt am Main, Germany}
\newcommand{\JLU}{Institut f\"ur Theoretische Physik,
Justus-Liebig-Universit\"at, Heinrich-Buff-Ring 16, 35392 Gie{\ss}en, Germany}
\newcommand{\HFHF}{Helmholtz Research Academy Hesse for FAIR (HFHF),
Campus Gie{\ss}en, 35392 Gie{\ss}en, Germany}
\newcommand{\ELTE}{Institute of Physics and Astronomy, ELTE E\"otv\"os Lor\'and University, P\'azm\'any P.\ s\'et\'any 1/A, H-1117 Budapest, Hungary}

\title{\boldmath Renormalization group invariant mean-field model for QCD at finite isospin density}

\author{Bastian B. Brandt}\affiliation{\BU}
\author{Volodymyr Chelnokov}\affiliation{\GU}
\author{Gergely~Endr\H{o}di}\affiliation{\ELTE}\affiliation{\BU}
\author{Gergely~Mark\'o}\affiliation{\BU}
\author{Daniel Scheid}\affiliation{\JLU}
\author{Lorenz von Smekal}\affiliation{\JLU}\affiliation{\HFHF}

\date{\today}

\begin{abstract}
QCD at nonzero isospin chemical potentials has phenomenological relevance for a series of physical systems and provides an ideal testground for the modeling of dense strongly interacting matter.
The two-flavor quark-meson model is known to effectively describe the condensation of charged pions in QCD that occurs in this setting.
In this paper, we derive a renormalization-group invariant mean-field formulation of the model and demonstrate that the resulting phase diagram and equation of state are in quantitative agreement with data from lattice QCD simulations
at small and intermediate isospin chemical potentials. In particular, the speed of sound from the model shows an excess over the conformal bound as previously seen in lattice computations in agreement with chiral perturbation theory. We then consider the speed of sound in the limit of large isospin chemical potentials and see that it approaches the conformal limit from above, in qualitative agreement with recent lattice results and in quantitative agreement with perturbation theory in the presence of a BCS gap. Finally, we consider the phase diagram in the approach to the chiral limit. We find that within the model the chiral phase transition connects to the pion condensation phase boundary in the chiral limit and we discuss the implications for the properties of the chiral transition point.
\end{abstract}

\maketitle

\section{Introduction}

The thermodynamics of strongly interacting matter is an important input for the phenomenological description of physical systems like the early universe, compact stars and their mergers, as well as heavy-ion collisions, which are used to study this kind of matter in the laboratory. Apart from the phase diagram, the equation of state (EoS) is the main input for the phenomenological modeling of such systems. While the common scenario for the evolution of the early universe does not include large quark densities, or chemical potentials in the grand canonical setting, except in the presence of large lepton flavor asymmetries during its early evolution~\cite{Wygas:2018otj,Middeldorf-Wygas:2020glx,Vovchenko:2020crk}, they are a key feature of compact stars and heavy-ion collisions, so that first principles knowledge about the EoS in the full parameter space of temperature $T$ and quark chemical potentials $\mu_f$ is needed. The dominant factor is typically a non-zero baryon density, but all of the systems mentioned above also include large charge, or, equivalently, isospin density components. 

The main tool to obtain non-perturbative first principles results for the phase diagram and the EoS are numerical simulations of lattice QCD. However, these suffer from the complex action problem at non-zero baryon chemical potential, prohibiting a direct calculation of the EoS at the parameters relevant for the above physical systems. In contrast, QCD at pure isospin chemical potential $\muI$, i.e., at vanishing baryon and strangeness chemical potentials, has a real action and can be simulated directly~\cite{Son:2000xc,Kogut:2002tm,Kogut:2002zg}. The associated phase diagram with physical quark masses has been mapped out a few years ago~\cite{Brandt:2017oyy}, including the phase of a Bose-Einstein condensate (BEC) of charged pions, and the EoS has been computed from direct simulations in the grand canonical setting~\cite{Brandt:2018bwq,Brandt:2022hwy}, as well as in a canonical setup at zero temperature~\cite{Detmold:2012wc,Abbott:2023coj,Abbott:2024vhj}.

Due to the complex action problem of lattice simulations at non-zero baryon chemical potential, model descriptions of the thermodynamics of strongly interacting matter have a long history. In particular, the phase diagram and the equation of state at non-zero isospin chemical potential have been studied within a multitude of models, including the Nambu-Jona-Lasinio (NJL) model~\cite{Toublan:2003tt,Frank:2003ve,Barducci:2004tt,He:2005sp,Zhang:2006gu,Sun:2007fc,Xia:2013caa,Zhang:2015baa,Brauner:2016lkh,Khunjua:2018sro,Khunjua:2018jmn,Avancini:2019ego,Lu:2019diy,Carlomagno:2021gcy,Lopes:2021tro,Liu:2021gsi,Liu:2023uxm,Carlomagno:2024xmi} and the linear sigma or quark-meson 
model~\cite{He:2005nk,Kovacs:2007sy,Herpay:2008uw,Kamikado:2012bt,Ueda:2013sia,Stiele:2013pma,Andersen:2018osr,Adhikari:2018cea,Folkestad:2018psc,Tawfik:2019tkp,Ayala:2023cnt,Tawfik:2023egf,Chiba:2023ftg,Ayala:2024sqm,Chiba:2024cny,Kojo:2024sca}. We also refer the reader to Ref.~\cite{Mannarelli:2019hgn} for a review.
Here we will focus on the quark-meson model. While the model cannot capture the physics of nuclear matter in QCD at finite baryon density, it provides an accurate description of the competing dynamics of chiral versus charged pion condensation at $\muI\neq0$ and can accurately reproduce the associated phase diagram, e.g.~\cite{Kamikado:2012bt,Adhikari:2018cea,Folkestad:2018psc}. The model also shows a crossover from the BEC phase with tightly bound pions to a BCS superconducting phase, where pions become weakly bound and take the form of pseudoscalar Cooper pairs~\cite{Son:2000xc}, which is still actively searched for in lattice studies~\cite{Brandt:2019hel,Cuteri:2021hiq}. This transition is analogous to the BEC/BCS crossover in ultracold fermionic quantum gases~\cite{Boettcher:2014xna} and happens in the model when the dynamically generated Dirac mass of the quarks rotates into a Majorana mass as $\muI$ increases. A similarly good agreement between the quark-meson model and the lattice data has been observed for a number of quantities related to the EoS~\cite{Ayala:2023cnt,Ayala:2024sqm}.

In this article, we perform a comprehensive comparison between the predictions of a renormalization-group (RG) invariant mean-field (MF) formulation \cite{Langfeld:1993zi,vonSmekal:1994zg} of the quark-meson model and recent lattice data for the phase diagram and the equation of state. 
The formulation should be equivalent to the one used in \cite{Andersen:2018osr,Adhikari:2018cea}, although in different regularization and renormalization schemes. Especially the latter seems to make a difference that is significant enough to briefly sketch the derivation of our  RG-invariant MF formulation here, for convenience. We then fix the RG-invariant scale parameter of the model by using the lattice data for the isospin density at $T=0$ within the BEC phase close to the phase boundary. Adjusting the mass-parameter of the scalar $\sigma$-meson to the chiral crossover transition temperature $T_{pc}(\muI=0)$ at zero isospin density, we then compare a range of other thermodynamic quantities. In particular, for the resulting speed of sound we find good agreement with lattice data and chiral perturbation theory \cite{Andersen:2023ofv} from the onset of pion condensation, surprisingly,  all the way up to the perturbative regime, where the asymptotic conformal limit $c_s^2=1/3$~\cite{Cherman:2009tw} is approached from above, and well within recent QCD predictions \cite{Fujimoto:2023mvc,Fukushima:2024gmp}. With the model parameters fixed, we furthermore compare the resulting finite temperature phase diagram to the one obtained on the lattice, and once more find very good qualitative and also good quantitative agreement for most of the phase boundaries. Finally, we scrutinize the behavior of the BEC phase boundary in the approach to the chiral limit,
where the model can serve as an effective theory 
to describe the competition between chiral symmetry breaking and pion condensation. In this regime lattice data is only available at half the physical pion mass, where the model shows good qualitative agreement for the change of the phase boundary with the results from lattice QCD. The model indicates that the pion condensation phase boundary connects to the chiral phase transition, showing that the chiral phase transition marks a multicritical point in the QCD phase diagram.

\section{Renormalization Group Invariant Mean-Field Potential}

Here we use the two-flavor quark-meson (QM) model as an effective theory to 
describe chiral symmetry breaking and charged pion condensation. The Lagrangian of the model with finite chemical potentials 
$\muq = (\mu_u +\mu_d)/2 $ and $\muI =  (\mu_u - \mu_d)/2 $ for average quark number and isospin imbalance reads,
\begin{equation}
 \begin{split}
  {\cal L}_{\rm QM} =&  \bar{\psi} \left( \slashed{\partial} + g
    (\sigma + i \gamma_5  
  \vec{\pi}\vec{\tau})- (\muq+\muI\tau_3 ) \gamma_0\right)  \psi \\
  & + \frac{1}{2} 
  (\partial_\alpha + 2
    \muI\delta^0_\alpha)\pi_+
    (\partial_\alpha - 2\muI
    \delta^0_\alpha)\pi_- \\
    & +\frac{1}{2}(\partial_{\alpha} \sigma)^2  + \frac{1}{2}(\partial_{\alpha}
 \pi_0)^2 + U(\rho^2,\pi^2)-c \sigma \, . ,
 \end{split}
 \end{equation}
 where $\pi_0 \equiv \pi_3 $, $\pi_\pm \equiv  \pi_1 \pm \iu \pi_2$, $\pi^2 \equiv \pi_+\pi_- = \pi_1^2+\pi_2^2$, and $\rho^2 = \sigma^2+\pi_0^2$. Note that with these definitions, the isospin charges of up and down quarks are $\pm 1$ while those of the charged pions are $\pm 2$, corresponding to the conserved isospin  current  
 \begin{equation}
 j^3_{\alpha} = \bar{\psi} \tau_3 \gamma_{\alpha} \psi 
+ \pi_- \partial_{\alpha}\pi_+ - \pi_+ \partial_{\alpha}\pi_- \, .
\end{equation}
At mean-field level, the quark contribution to the Landau free-energy density at temperature $T$ is given by \cite{Kamikado:2012bt}
\begin{equation}
\begin{split}
    \Omega_q(T,\muq,\muI) &= \\
    &\hspace{-1cm}
    -2 T N_c \!\idp \sum_\pm \Big( \ln\cosh\frac{E_\pm -\muq}{2T} \\
    & \hspace{2.4cm} +  \ln\cosh\frac{E_\pm +\muq}{2T} \Big)\, ,
    \end{split} 
\end{equation}
where, $E_\pm^2 = g^2 \pi^2 + (\epsilon_p \pm \muI)^2 $, and $\epsilon_p^2 = p^2 + g^2 \rho^2 $, and $N_c = 3$ is used for the number of colors throughout. 
The complete mean-field expression includes the mesonic effective potential $U(\rho^2,\pi^2) $ together with explicit symmetry breaking and charged-pion seagull terms, and reads
\begin{equation}
\begin{split}
    \Omega(T,\muq,\muI) 
    =
    U(\rho^2,\pi^2)  -c \sigma - 2\muI^2 \pi^2 
    + \Omega_q(T,\muq,\muI) \,.  
    \end{split}\label{eq:mfOmega}
\end{equation}
Here, we use the simplest  $O(4)$-symmetric potential for the scalar and pseudoscalar meson fields with quartic self-interactions, 
\begin{equation}
 U(\rho^2,\pi^2) = \frac{m^2}{2} (\rho^2+ \pi^2) +\frac{\lambda}{4} (\rho^2+\pi^2)^2   \,. \label{eq:baresymmPot} 
\end{equation}

As usual, the zero-temperature contribution from the quarks contains ultraviolet (UV) divergent vacuum terms. Unlike the case of a pure baryon or symmetric quark chemical potential $\muq$, inside the pion condensation phase (for $\pi^2\not = 0$) there are UV divergent vacuum terms proportional to $ \pi^2\muI^2 $ \cite{Andersen:2018osr}. 
To see this explicitly, we define 
\begin{equation}
\Omega^0_q(\muI^2) \equiv \Omega_q(T=0,\muq=0,\muI)\,,    
\end{equation}
and, with $E_\pm \equiv E_\pm(p^2, g^2 \rho^2, g^2\pi^2, \muI) $, write
\begin{equation}
    \Omega_q^0(\muI^2)  = -\frac{N_c}{\pi^2 } \sum_\pm \int p^2 dp \; E_\pm(p^2, g^2 \rho^2, g^2\pi^2, \muI) \, ,
\end{equation}
which is even in $\muI$ because  $E_-(\muI) = E_+(-\muI)$.
Taylor expanding around $\muI = 0 $, one observes that the first two terms are ultraviolet divergent,
\begin{equation}
\begin{split}
    \Omega_q^0(\muI^2) &= -\frac{N_c}{\pi^2 } \int p^2 dp \, \bigg(
    2 \sqrt{p^2 + g^2 (\rho^2+\pi^2)}\\
    &\hspace{.5cm}+ \frac{g^2 \pi^2 \muI^2}{\sqrt{p^2 + g^2 (\rho^2+\pi^2)}^3} \bigg) \\
    &\hspace{.5cm}+ \mbox{UV finite } \mathcal O(\muI^4)\, . \end{split}
    \label{eq:TaylorVacmu}
\end{equation}
When we subtract the logarithmic divergence from the second term, proportional to $\pi^2 \muI^2$, we must be careful not to change the Silver-Blaze behavior by the subtraction. Note that the unsubtracted original $\Omega_q^0(\muI^2)$ is actually independent of $\muI $ as long as $g^2\pi^2= 0$, i.e.~in the vacuum, $\mu_I<\mu_\mathrm{c}=m_\pi/2$.\footnote{Note that the onset of pion condensation is at half the pion mass with the convention for isospin chemical potential used here. In Fig.~\ref{fig:SoundSpeed_compare_3}, however, we compare to results obtained in a different convention, which results in a scaled $\muI$ axis.} 
Because the subtracted Landau free-energy density must vanish together with all its $\muI$-derivatives as long as $\muI< \mu_\mathrm{c} $, it is of course not analytic at this point, which is the bifurcation point in the zero-temperature gap equation for the pion condensate.

In order to make the $O(4)$ symmetry for $\muI=0$ explicit, we first introduce the following new field variables which furthermore turn out to be the renormalization group invariant and hence physical variables in our RG-invariant mean-field formulation, 
\begin{equation}
    M^2 \equiv g^2 (\rho^2+ \pi^2)\, , \;\; \mbox{and} \;\; \Delta^2 \equiv g^2 \pi^2\; .
\end{equation}
Because of the two independent logarithmic UV divergences in Eq.~\eqref{eq:TaylorVacmu}, proportional $M^4 $ and to $\muI^2 \Delta^2$, respectively, we need two independent renormalization constants. One is the usual mean-field renormalization constant of the quartic meson self-interactions, defined by $\lambda_R = Z_\lambda^{-1} \lambda$, and the other renormalizes the   quark-meson Yukawa coupling, via $g_R^2 = Z_g^{-1} g^2$, where the subscript $R$ denotes the multiplicatively renormalized (and hence renormalization scale $\muRG $ dependent but UV finite) quantities. The latter in particular also implies that one needs meson field $\phi =(\sigma,\vec\pi) $ and mass renormalization, introducing $\phi_R = Z_\phi^{-1/2} \phi $ and $m_R^2 = Z_m^{-1} m^2$.
These are not independent, however, but determined from $Z_g Z_\phi = 1$ and $Z_m Z_\phi  = 1$, so that the meson  field variables $M$ and $\Delta$ are renormalization group invariant, and so is the ratio of the mass parameter in the meson Lagrangian and the Yukawa coupling,  $ m^2_R/g^2_R = m^2/g^2$.

Regularization and renormalization follow standard procedures \cite{Skokov:2010sf,Andersen:2018osr,Folkestad:2018psc,Ayala:2023cnt} as summarized for convenience in the appendix. Since all quantities below are renormalized and finite, or RG invariant, we drop their subscript $R$ again, from now on. For our renormalization group invariant formulation it is convenient to introduce two new couplings $u\equiv g^4/\lambda $ and $ v = g^2$, with $Z_u  = Z_g^2/Z_\lambda $ and $Z_v = Z_g$, and corresponding $\beta$ functions obtained from their RG scale $\muRG$ dependences, which  at the mean-field level are given by 
\begin{align}
    \beta_u(u) &\equiv \muRG \frac{d u(\muRG) }{d\muRG} = u \frac{d Z^{-1}_u }{d\ln\muRG}=  \frac{N_c}{\pi^2}  \, u^2 \,  , \label{eq:betau}\\
    \beta_v(v) &\equiv \muRG \frac{d v(\muRG) }{d\muRG} = v \frac{d Z^{-1}_v }{d\ln\muRG} = \frac{N_c}{2\pi^2} \, v^2 \, . \label{eq:betav}
\end{align}
The first equation is used to define the RG invariant scale parameter $M_0$ of the mean-field model \cite{Langfeld:1993zi} in terms of general $\muRG$ and $u \equiv u(\muRG)$ (but independent of $v$),
\begin{align}
    M_0 = \muRG \exp\Big\{ -\int_{u_0}^u \frac{dl}{\beta_u(l)} \Big\}\, , \;\;\mbox{with}\;\;   u_0 = u(\muRG_0) \, .
\end{align}
As our first renormalization condition, we use the initial condition $\muRG_0 =M_0$ 
here such that the reference scale $\muRG_0$ is fixed to the RG invariant constituent quark mass in the vacuum, 
\begin{align*}
    M_0 \equiv g_0\sigma_0 \, , \;\mbox{where}\;\;   g_0 = g(\muRG_0)\; \mbox{and} \;\; \sigma_0 = \sigma(\muRG_0) = f_\pi\, ,
\end{align*}
where $f_{\pi} \approx 93$~MeV is the usual  PCAC pion decay constant. 
The second renormalization condition that we need to introduce is to fix the residue of the pole in the (RG scale $\muRG $ dependent) meson propagator in the vacuum to be unity at the reference scale $\muRG_0$, which for the charged pion two-point function $\Gamma^{(2)}_{\pi_\pm}(p,\muI)$  at vanishing four-momentum, $p=(p_0,\vec p) = 0$, for $T=\muq=0 $ and
$\mu_I \to m_\pi/2$ from below, implies that
\begin{equation}
    \Gamma^{(2)}_{\pi_\pm}(0,\muI) = \frac{g^2(\muRG)}{g_0^2} \, \big( m_\pi^2 - (2\muI)^2 \big)\,.\label{eq:rencond_pion}
\end{equation}
The reason why a wave function renormalization appears for the charged pions despite having no momentum dependent divergence, is because the chemical potential essentially enters as a shift to the frequency at zero temperature. For a general discussion based on a scalar model see \cite{Marko:2014hea}.

Our full effective potential is then explicitly verified to be renormalization group invariant, satisfying the simple renormalization group equation
\begin{align}
    \Big(\muRG\frac{\partial} {\partial\muRG } + \beta_u(u) \frac{\partial} {\partial u }+ \beta_v(v) \frac{\partial} {\partial v } \Big) \, \Omega(T,\muq,\muI) &= 0\, . 
\end{align}
It is therefore possible to eliminate the explicit RG scale dependence of the renormalized vacuum contributions together with the scale-dependent couplings $u$ and $v$ in favor of the single scale parameter $M_0$ and the couplings $u_0$ and $v_0$ defined at this reference scale. The latter are then implicitly determined by our two renormalization conditions which allow them to be eliminated as well. Therefore, the full effective potential, as function of the RG invariant field variables $M$, $\Delta$ and $\Sigma \equiv g \sigma $, can be expressed in terms of physical parameters and our single free scale parameter $M_0$ -- the analogue of $\Lambda_\mathrm{QCD}$ in the model. The result is 
\begin{widetext}
\begin{align}
    \Omega(T,\muq,\muI) &=  \frac{1}{2} \, \Big(\frac{(3m_\pi^2 - m_\sigma^2)f_\pi^2}{2M_0^2}  -\frac{N_c}{2\pi^2} \,M_0^2 \Big)    M^2  +\frac{1}{4} \,
     \Big(
     \frac{(m_\sigma^2- m_\pi^2)f_\pi^2}{2M_0^4}
     + \frac{3N_c}{4\pi^2}  \Big) M^4  - \frac{m_\pi^2f_\pi^2}{M_0} \, \Sigma 
    \nonumber\\[4pt]
     &   \hskip -1.2cm 
    -  \frac{2 f_\pi^2}{M_0^2} \, \muI^2  \Delta^2
     - \frac{N_c}{4\pi^2} \, M^4  \ln\big(M/M_0\big)  + \frac{N_c}{\pi^2} \, \muI^2 \Delta^2 \Big(\ln\big(M/M_0\big)- F^0_q\big(m_\pi/(2M_0)\big)\Big)  + R^0_q(M^2,\Delta^2;\muI^2) \label{eq:mfOmegaRGphys}\\[6pt]
     & \hskip -1.2cm - 2 T N_c \idp \sum_\pm \bigg( \ln\Big( 1+ e^{-\frac{|E_\pm -\muq| }{T}} \Big) +\ln\Big( 1+ e^{-\frac{|E_\pm +\muq| }{T}} \Big) \bigg) 
 - 2 N_c \idp \sum_\pm \theta\big(|\muq| - E_\pm \big) \big( |\muq| - E_\pm \big) \, ,
    \nonumber 
\end{align}
\end{widetext}
as derived in the appendix. We have been a bit sloppy not distinguishing the effective potential from the thermodynamic grand potential here, where the latter is of course obtained upon minimizing the former with respect to the two independent field variables $M$ (or $\Sigma$) and $\Delta$, dependent on temperature and chemical potentials.

We have furthermore introduced the subtracted and UV finite residual of the field dependent vacuum terms,
\begin{align}
    R^0_q(M^2,\Delta^2;\muI^2) &\equiv \label{eq:VacuumResidual}\\
    &\hskip -.4cm \Omega_q^0(\muI^2) -   \Omega_q^0(0) -
    \muI^2 \frac{\partial }{\partial \muI^2 } \Omega_q^0(\muI^2)\Big\vert_{\muI^2=0} \, .\nonumber
\end{align}
Importantly,  for $\Delta = 0 $, this residual vacuum contribution still contains the degeneracy pressure of a Fermi gas at finite isospin chemical potential, when $\muI^2  > M^2 = g^2 \rho^2 $. This ground-state pressure $G_q^0(M;\muI) $  is not affected by the subtraction in (\ref{eq:VacuumResidual}) and can be isolated,
\begin{align}
    R^0_q(M^2,\Delta^2=0;\muI^2) &\equiv 
    - G^0_q(M;\muI) \, ,
  \end{align}
where
  \begin{align}
     G^0_q(M;\muI) &=   4 N_c \idp \theta\big(|\muI| - \epsilon_p \big) \big( |\muI| - \epsilon_p \big) \, ,  \nonumber
\end{align}
  with $\epsilon_p = \sqrt{p^2 + M^2}$.
Since $G^0_q(M;\muI) \equiv 0 $ for $\muI^2 < M^2$, in this regime, we can then use the expansion,
\begin{align}
        R^0_q(M^2,\Delta^2;\muI^2) &= \frac{N_c}{\pi^2} \, \Delta^2\muI^2 \, \Big( F_q^0(\muI/M)  + \mathcal O (\Delta^2/M^2) \Big)  , \nonumber
\end{align}
with a transcendental function $F^0_q(x) $ given explicitly in the appendix.

Moreover, the value of the leading contribution of $\mathcal O (\Delta^2) $ to $  R^0_q(M^2,\Delta^2;\muI^2)$ at $\muI = m_\pi/2$ is also subtracted, in the second line of Eq.~(\ref{eq:mfOmegaRGphys}), due to our second renormalization condition for the charged pion two-point function in Eq.~(\ref{eq:rencond_pion}).
Compared to minimal subtraction (or $\overline{MS}$) this amounts to subtracting an additional small but finite contribution, of relative size 
$C_\phi = - F^0_q\big(m_\pi/(2M_0)\big) \approx 0.013$ for $m_\pi = 139 $~MeV and $M_0 = 350$~MeV, for example, in the meson field renormalization constant, see the appendix. Otherwise, the RG-scale dependent prefactor $Z(\muRG ) = g^2(\muRG)/g_0^2 $ of the vacuum pion propagator   \eqref{eq:rencond_pion} for $\mu_I \to m_\pi/2$ from below would not be unity at the reference scale but some  $Z_0 \equiv Z(\muRG_0) $, given by 
\[
Z_0  = 1 - \frac{N_c g_0^2}{2\pi^2} \, \Big( F_q^0\big(m_\pi/(2 \muRG_0)\big)  + C_\phi \Big) \, .
\]
One would then need to rescale the physical pion mass by $ m_\pi^2 \to  Z_0 m_\pi^2 $, accordingly, to compensate this finite multiplicative renormalization factor in its definition. This would automatically readjust the onset of pion condensation from $\mu_I = Z^{-1/2}_0 m_\pi/2 $ back to half the physical $m_\pi/2$ at the same time, precisely as one expects from the meson-field renormalization constant $Z_\phi = Z_g^{-1}$ with $Z_g = Z_v $ given in Eq.~\eqref{eq:Zv} of the appendix. 

The RG-invariant gap equations for $M$ and $\Delta $, here  at $T=\muq=0 $ for simplicity, are then obtained straightforwardly from Eq.~\eqref{eq:mfOmegaRGphys} as follows
\begin{widetext}
\begin{align}
    0 &= \left(\frac{(3m_\pi^2 - m_\sigma^2)f_\pi^2}{2M_0^2}  -\frac{N_c}{2\pi^2} \,M_0^2 \right)    M  + \,\left( \frac{m_\pi^2 f_\pi^2}{M_0^4} - \frac{(3m_\pi^2 - m_\sigma^2)f_\pi^2}{2M_0^4}  + \frac{3N_c}{4\pi^2}  \right) M^3  - \frac{m_\pi^2f_\pi^2}{M_0} \, \frac{M}{\sqrt{M^2-\Delta^2}} 
    \nonumber\\[4pt]
     &   \hskip 1cm 
     - \frac{N_c}{\pi^2} \, M^3 \Big( \ln\big(M/M_0\big)  + \frac{1}{4}\Big)  + \frac{N_c}{\pi^2} \, \frac{\muI^2 \Delta^2}{M} + 2 M \frac{\partial}{\partial M^2} R^0_q(M^2,\Delta^2;\muI^2) \, , \label{eq:gapM}\\[6pt]
0 &= \frac{m_\pi^2f_\pi^2}{M_0} \, \frac{\Delta}{\sqrt{M^2-\Delta^2}}  -  \frac{4 f_\pi^2}{M_0^2} \, \muI^2  \, \Delta     
     + \frac{2N_c}{\pi^2} \, \muI^2 \, \Delta \Big( \ln\big(M/M_0\big) + C_\phi \Big) + 2 \Delta \frac{\partial}{\partial \Delta^2} R^0_q(M^2,\Delta^2;\muI^2) \, . \label{eq:gapDelta}
\end{align}
\end{widetext}
The solution for $\Delta = 0 $, i.e.~the trivial solution to Eq.~\eqref{eq:gapDelta},  reduces  to $M=M_0$ from  Eq.~\eqref{eq:gapM}, as it must.

For the bifurcation analysis of the $\Delta$ gap equation, we linearize the non-trivial solution to Eq.~\eqref{eq:gapDelta} as
\begin{align}
       0 =&\, \frac{m_\pi^2f_\pi^2}{M_0 M} \, \Big(1+ \mathcal O\big(\Delta^2/M^2\big)\Big) -  \frac{f_\pi^2}{M_0^2} \, (2\muI)^2 \nonumber \\
     &+ \frac{2N_c}{\pi^2} \, \muI^2 \, \ln\big(M/M_0\big) 
     +  \frac{2N_c}{\pi^2} \, \muI^2 \Big( F^0_q(\muI/M) \label{eq:bifurcation_point}\\
     &\hskip 1cm- F^0_q\big(m_\pi/(2M_0)\big) + \mathcal O\big(\Delta^2/M^2\big) \Big)\, , \nonumber
\end{align}
which confirms that for  $\Delta=0$  in the vacuum, where $M=M_0 $, the bifurcation point, i.e.~the non-trivial root of the gap equation, occurs at $2\muI= m_\pi$ as it must. The effective potential is of course non-analytic at this point, in fact, one has $R^0_q \equiv 0$ in the vacuum (for the trivial solution $\Delta=0$ below the bifurcation point). In our renormalization scheme, we explicitly see that for $\mu_I \to m_\pi/2$ from below
only the first line in  Eq.~(\ref{eq:bifurcation_point}) remains, which thus agrees with Eq.~(\ref{eq:rencond_pion}), up to the factor of $g^2$ from the derivatives of $\Delta$ with respect to $\pi $ and with $g_0^2 = v_0 = M_0^2/f_\pi^2$ at the reference scale.      

As a final remark on our RG-invariant mean-field formula for the two-flavor quark-meson model in Eq.~(\ref{eq:mfOmegaRGphys}), we note that it is consistent with QCD inequalities \cite{Cohen:2003ut} for the Landau free energy densities
\begin{align}
   \Omega_B(T,\muq) &\equiv    \Omega(T,\muq,\muI=0) \, ,\\
   \Omega_I(T,\muI) &\equiv    \Omega(T,\muq=0,\muI) \, ,
\end{align}
with chemical potentials either only for baryon density (where $\mu_B = 3 \muq$) or only for isospin density (with $\muI$ for isospin charges $\pm 1$ of up and down quarks here), respectively. For these one then readily verifies from Eq.~(\ref{eq:mfOmegaRGphys}) that
\begin{equation}
    \Omega_B(T,\mu) \ge  \Omega_I(T,\mu) \, . \label{eq:InequalityQCD}
\end{equation}
To show this, one first notes that outside the pion-condensation phase, 
this inequality in (\ref{eq:InequalityQCD}) is in fact saturated
in our mean-field model, with 
\begin{equation}
    \Omega_B(T,\mu) =  \Omega_I(T,\mu) \; , \;\; \mbox{for} \;\; \Delta = 0 \, .
\end{equation}
The inequality in the general case then follows from observing that a non-vanishing pion condensate produces additional contributions to $ \Omega_I(T,\mu) $ that are negative definite and absent from  $ \Omega_B(T,\mu)$. 

\section{Comparison with Lattice Data}

Apart from the physical parameters $m_\pi $, $f_\pi$ and $m_\sigma $, there is only the single parameter  $ M_0 $ in the RG-invariant mean-field theory that sets the overall scale of all dimensionful quantities. As the results, in units of $M_0$, will only depend on $m_\pi/M_0 $, $f_\pi/M_0$ and $m_\sigma/M_0 $, in principle, we could choose it to be anything we like. 

\begin{figure}
    \centering
    \includegraphics[width=\linewidth]{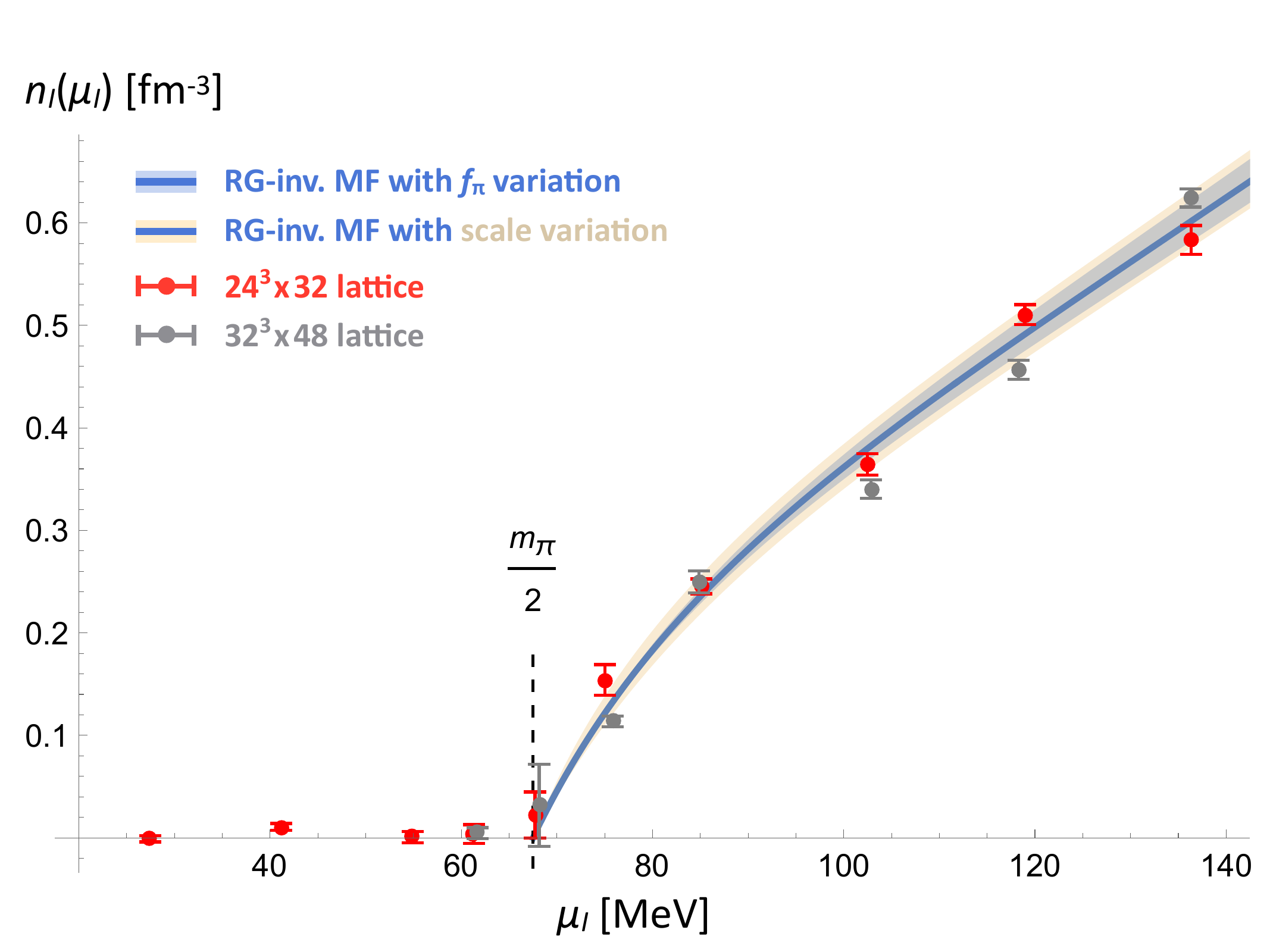}
        \caption{Zero-temperature isospin density in physical units compared to the data from lattices of size $24^3 \times 32 $ (red) and $32^3 \times 48 $ (gray), with spacings $a= 0.22$~fm and $a=0.15$~fm and $m_\pi \simeq 135$~MeV~\cite{Brandt:2022hwy}, using $M_0 = 350$~MeV and $f_\pi = 90$~MeV for the best overall description (solid blue line); the light blue band shows the effect of varying $f_\pi$ between 88~MeV and 92~MeV, and the light yellow band that of varying the scale parameter $M_0$ between 300~MeV and 400~MeV at fixed $f_\pi$. }
    \label{fig:IsospinDensityData}
\end{figure}

In physical units, however, the lattice data for the isospin density at zero temperature from Ref.~\cite{Brandt:2022hwy} can be used to constrain $M_0 $. To do this, we first fix $m_\pi = 135$~MeV, which is the pion mass used in the simulations of Ref.~\cite{Brandt:2022hwy} for two different lattices of sizes $24^3 \times 32 $ and $32^3 \times 48  $ and lattice spacings of $a= 0.22$~fm and $a=0.15$~fm. 
This leaves us with $f_\pi $ and $m_\sigma $ as adjustable parameters. 
The value of $m_\sigma $ is particularly poorly constrained. In fact, one might argue that it is not a very well defined physical parameter in the first place, remembering that the $\sigma $-meson is best described as a broad two-pion resonance corresponding to a pole
at $m_\sigma\simeq ( 450 - \iu\, 275 )$~MeV on the unphysical Riemann sheet \cite{Pelaez:2015qba}.  
Luckily, the RG-invariant mean-field results for the isospin density at zero temperature turn out to be practically insensitive to the $\sigma$-meson mass parameter $m_\sigma $, with no noticeable differences at all, when choosing any value of $m_\sigma$ somewhere in a range between about $300$ and $700$~MeV, so the results should also remain unchanged when averaging over a correspondingly broad $\sigma $ spectral function.  
For different values of $M_0$ we are then left with adjusting the pion decay constant $f_\pi $  to match the lattice data. Using a simple reduced $\chi$-squared criterion for the goodness of the fits we obtain the best overall description of the lattice data for values of $M_0 $ between 340 and 360~MeV. This nicely concurs  with the typical scale used in many previous quark-meson model studies and its interpretation as the constituent quark mass in the vacuum. Thus using $M_0 = 350$~MeV (and, for completeness, $m_\sigma = 470 $~MeV, here), the result for $f_\pi = 90$~MeV, shown as the solid blue line in Fig.~\ref{fig:IsospinDensityData}, describes the data fairly well. It can in fact be seen as providing a global fit to both data sets. Moreover, the narrow light-blue band shows the effect of varying $f_\pi $ between 88~MeV (lower edge) and 92~MeV (upper edge). The somewhat wider light-yellow band in Fig.~\ref{fig:IsospinDensityData}, on the other hand, is obtained from varying the RG-invariant scale parameter $M_0$ of the model between 300~MeV and 400~MeV, i.e.~by $\pm 50$~MeV around its central value, while keeping 
$f_\pi = 89$~MeV fixed.
This wider band covers the $f_\pi$ variations above, and it intersects with the error bars of all data points. In the figures below, we will therefore use this more conservative wider band as an indication of the present uncertainty in fixing the model parameters to the available data, in the range between the onset of pion condensation and $\muI \lesssim m_\pi $ as shown in Fig.~\ref{fig:IsospinDensityData}.

\begin{figure}
    \centering
    \includegraphics[width=\linewidth]{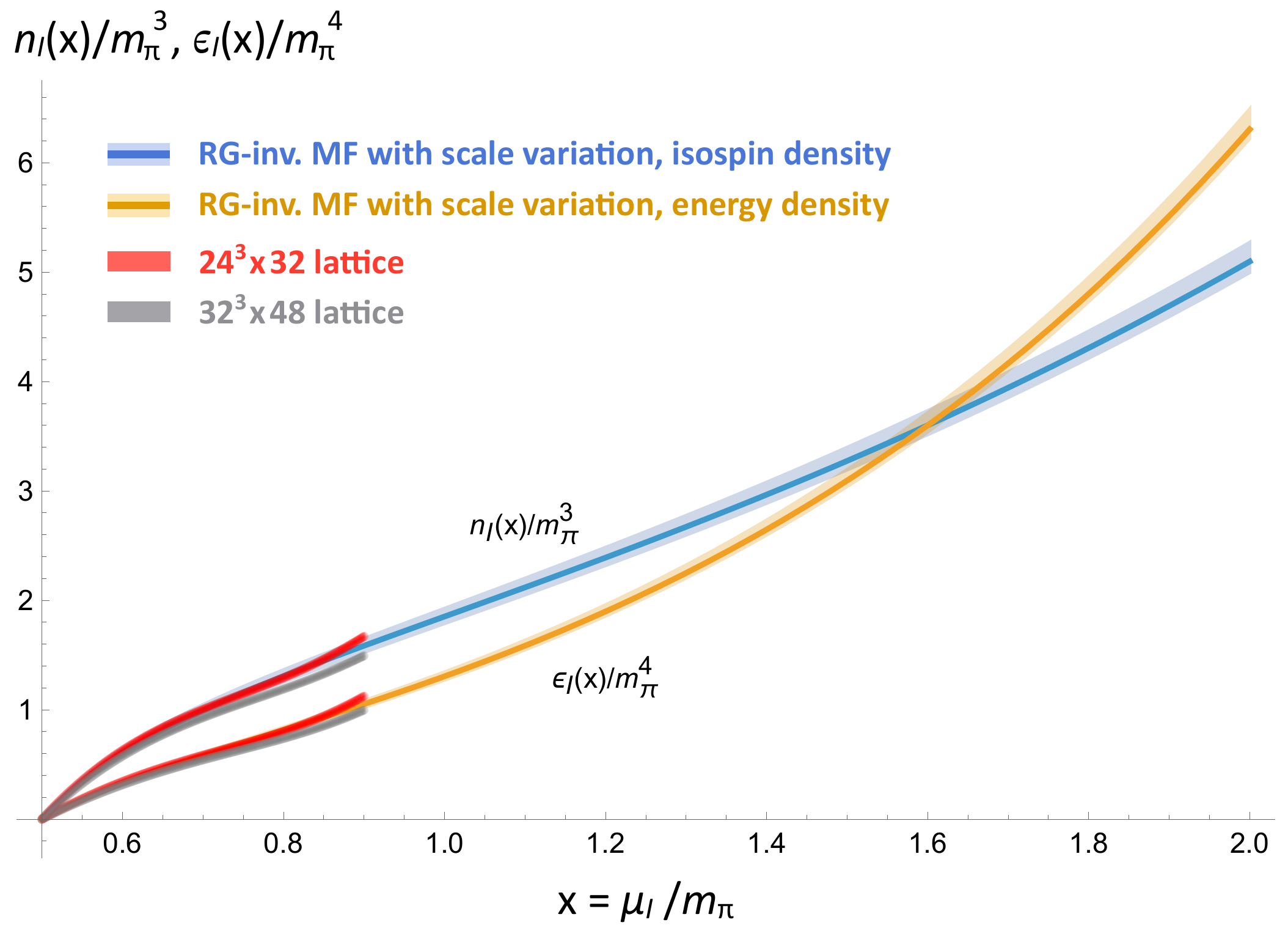}
    \caption{Isospin (upper curves) and energy densities (lower curves) in units of the pion mass as obtained from the quark-meson model (solid blue and yellow lines with bands indicating the effects of varying the RG-invariant scale parameter $M_0$ by $\pm 50$~MeV, see text)  and in lattice QCD~\cite{Brandt:2022hwy}. The color code for the latter is the same as in Fig.~\ref{fig:IsospinDensityData}.}
    \label{fig:IsospinandEnergyDensity}
\end{figure}

\begin{figure}[b]
    \centering
    \includegraphics[width=\linewidth]{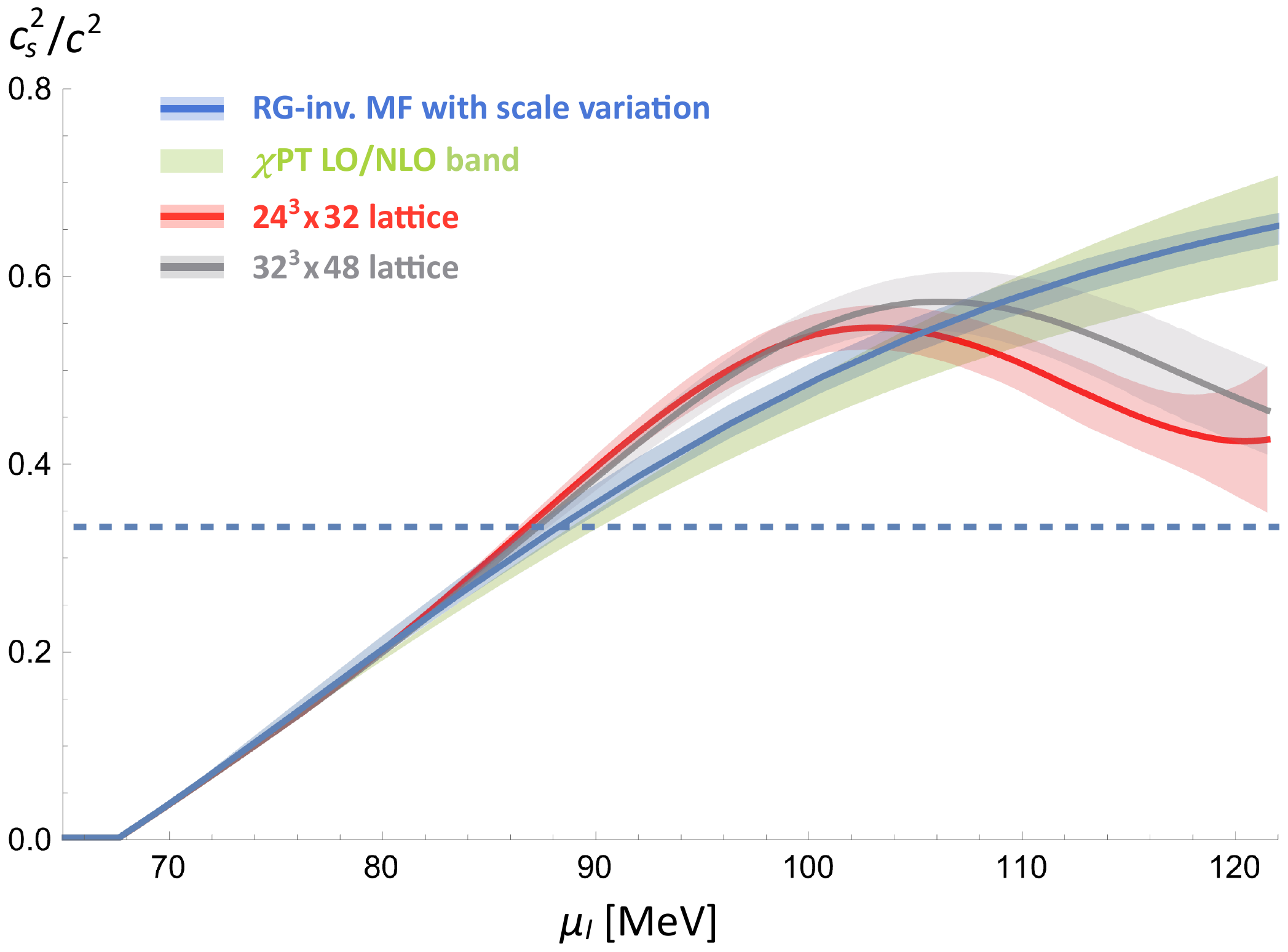}
    \caption{Results for the speed of sound as obtained in the quark-meson model and on the lattice~\cite{Brandt:2022hwy}. The color code for these results is the same as in Fig.~\ref{fig:IsospinDensityData}. In addition, we also show the result from chiral perturbation theory from Ref.~\cite{Andersen:2023ofv} as light-green band, as described in the text.}
    \label{fig:SoundSpeed_compare}
\end{figure}

Using the central values of $M_0=350$~MeV and $f_\pi = 90$~MeV, together with the same band of $\pm 50$~MeV variations in the scale parameter $M_0$, the isospin density and the energy density %$M_0 $ 
in units of their respective appropriate powers of $m_\pi$ are shown in Fig.~\ref{fig:IsospinandEnergyDensity} for isospin chemical potentials from the onset of pion condensation at $\muI = m_\pi/2$ up to $\muI = 2 m_\pi $ together with the interpolating bands for the two lattice data sets from Ref.~\cite{Brandt:2022hwy}.\footnote{These units are convenient to plot isospin $n_I$ and energy density $\epsilon_I$ on the same scale which is not possible when plotting, e.g., $n_I(\mu_I) $ in units of fm$^{-3}$ and $\epsilon_I(\mu_I) $ in units of MeV  fm$^{-3}$.} The resulting speed of sound is compared in Fig.~\ref{fig:SoundSpeed_compare} to the corresponding  interpolation bands for the lattice data from Ref.~\cite{Brandt:2022hwy} and chiral perturbation theory ($\chi$PT), as here represented by the light-green band between the leading order curve (LO, upper edge) and the next-to-leading order result (NLO, lower edge) from Ref.~\cite{Andersen:2023ofv}. 
In this range,  the RG-invariant  quark-meson model calculation is consistent with $\chi$PT and both are in good qualitative agreement with the lattice data. The effective theories show less small-scale features than the interpolating bands of the lattice data, which also shows in the fact that the speed of sound develops a peak at larger isospin chemical potentials, as we discuss next.

\begin{figure}
    \centering
     \includegraphics[width=\linewidth]{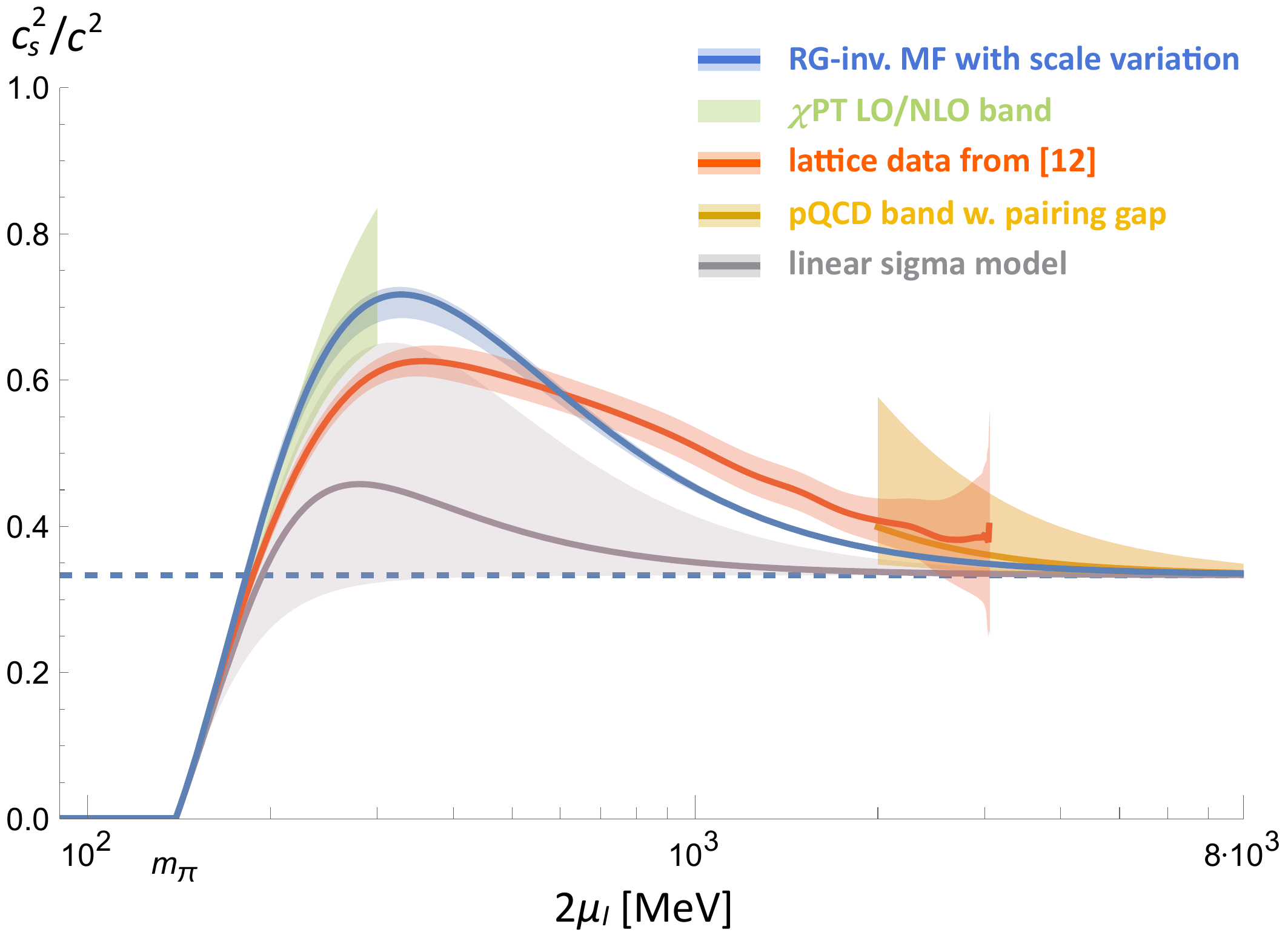}
    \caption{The speed of sound (solid blue) compared to the lattice data from Ref.~\cite{Abbott:2024vhj} (red line with light red error band), here using
    physical values for pion mass and decay constant ($m_\pi = 139.57 $~MeV for the onset of pion condensation, and $f_\pi = 92$~MeV, other parameters unchanged).
    Also shown are the $\chi$PT result from Ref.~\cite{Andersen:2023ofv} (light green band between LO and NLO results as upper and lower edges), and the most recent perturbative estimate with contributions from the superconducting pairing gap included from \cite{Fukushima:2024gmp} (yellow line with light yellow error band). The gray line and light-gray band show a family of analytic results, from Eq.~\eqref{eq:cslsm} for the linear sigma model for comparison (see text).
    }
    \label{fig:SoundSpeed_compare_3}
\end{figure}

This feature of a speed of sound peak at larger chemical potentials is shown in  Fig.~\ref{fig:SoundSpeed_compare_3} where its position (beyond the range of applicability of $\chi$PT) agrees with the lattice results obtained in the canonical setting of Ref.~\cite{Abbott:2024vhj}. For the comparison with this data (red line with light red error band), we have slightly adapted the pion mass and decay constant in our RG-invariant mean-field calculation (blue line) to concur with the ones used for the simulations of Ref.~\cite{Abbott:2024vhj}, changing from $m_\pi=135$~MeV and $f_\pi = 90$~MeV to $m_\pi=139.57 $~MeV and $f_\pi = 92$~MeV, while keeping the scale parameter fixed at $M_0=350$~MeV together with the same $\pm 50$~MeV (light blue) variation band, as used to describe the lattice data of Ref.~\cite{Brandt:2022hwy} in the figures above. Additionally shown in Fig.~\ref{fig:SoundSpeed_compare_3} are the recent  perturbative estimates including the superconducting pairing gap from Ref.~\cite{Fukushima:2024gmp} (light yellow band), and the chiral perturbation theory result from Ref.~\cite{Andersen:2023ofv}, as already included in Fig.~\ref{fig:SoundSpeed_compare} as well.

For another comparison, we have  also included in Fig.~\ref{fig:SoundSpeed_compare_3} a family of analytic results from the linear sigma model (gray line and band), where one readily works out from the mean-field equations of state \cite{Andersen:2010vu,Kamikado:2012bt,Scheid:2023} that
\begin{equation}
\frac{c_s^2}{c^2} =  \frac{1 - y^2 + x^4 (y^2-3) + 2 x^6}{3 (y^2 -1) + x^4 (y^2-3) + 6 x^6 } \, , \label{eq:cslsm}
\end{equation}
for $x=2 \muI/m_\pi \ge 1 $, and with $y= m_\sigma/m_\pi$. Of course, this expression reduces to the corresponding non-linear sigma model result  $c^2_s/c^2 = (x^4-1 )/(x^4+3) $ in the limit $y\to\infty$, i.e.~for $m_s\gg m_\pi $. This then yields the upper (LO $\chi$PT) edge of the green $\chi$PT band which always approaches $c_s^2/ c^2 \to 1 $ for asymptotically large chemical potentials. Although the qualitative effect of the finite $m_\sigma $, to restore the conformal limit $c_s^2/ c^2 \to 1/3 $, is thus quite significant, in stark contrast to our RG-invariant quark-meson model calculation, the linear sigma model result is very sensitive to this $\sigma$-meson mass parameter. To demonstrate this we have plotted the linear sigma model speed of sound \eqref{eq:cslsm} in Fig.~\ref{fig:SoundSpeed_compare_3} with the same $m_\pi$
and $m_\sigma=470$~MeV (solid gray line). The light gray band has then been
obtained from varying $m_\sigma \in [235,940]$~MeV, i.e., up and down by a factor of two relative to the central value, indicating the dependence of this result on variations of the $\sigma$-meson mass in a range where the RG-invariant quark-meson model result remains practically independent of this parameter as explained above.

These plots highlight how surprisingly well the RG-invariant quark-meson model calculation is able to describe the presently available QCD knowledge, including lattice data, chiral perturbation theory and perturbative estimates, over a remarkably wide range of almost two orders of magnitude in the isospin chemical potential, from the onset of pion condensation up to the perturbative regime and its approach to the conformal limit.

\begin{figure}
    \centering
    \includegraphics[width=\linewidth]{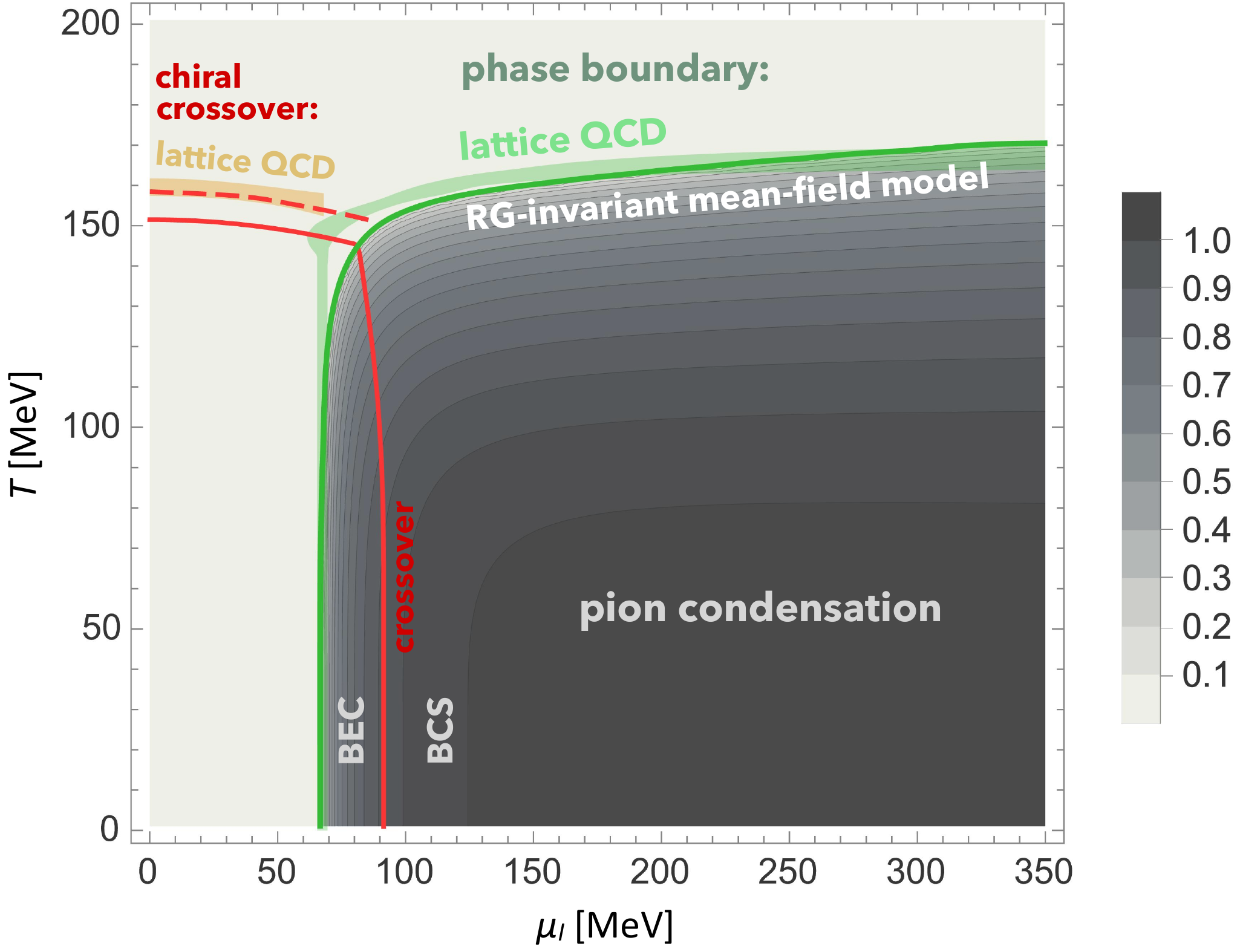}
    \caption{Phase diagram in the $(\muI,T)$-plane from the RG-invariant quark meson model calculation (with $M_0=350$~MeV, $m_\pi=135$~MeV, $f_\pi = 90$~MeV and $m_\sigma=470$~MeV): chiral crossover (dashed red), half-value  of the chiral order parameter $\Sigma$ (solid red), and  
    contour plot of the normalized pion condensate $\Delta/M_0$ with second-order phase boundary (solid green), compared to the chiral crossover (light orange band) and pion condensation phase boundary (light green band), as
    determined on the lattice in  Ref.~\cite{Brandt:2017oyy}.}
    \label{fig:muI_Y_Delta}
\end{figure}

\section{Finite Temperature Phase Diagram and Chiral Limit}

After this comprehensive comparison of the zero-tempe\-rature results, we finally turn to the phase diagram in the plane of isospin chemical potential $\muI$ and temperature $T$ in Fig.~\ref{fig:muI_Y_Delta}. The input parameters here are all fixed to the lattice data in Fig.~\ref{fig:IsospinDensityData} as discussed above, with  $m_\pi=135$~MeV, $f_\pi = 90$~MeV, $M_0 = 350$~MeV, and  $m_\sigma=470$~MeV. As also mentioned above already, $m_\sigma $ is thereby not constrained by the zero-temperature data but adjusted, so that the pseudo-critical temperature of the chiral transition at $\muI = 0 $ (in both cases here defined by the inflection point in the order parameter)  agrees with that determined on the lattice in Ref.~\cite{Brandt:2017oyy}. As a first consistency check, the curvature of the chiral transition (dashed red) then also agrees with the lattice determination (light orange band) for all $\muI$ up to the phase boundary of pion condensation. The chiral condensate develops a kink at the phase boundary to the pion condensation phase, so that there is no such inflection point inside the pion condensation phase. We have therefore also included the half-value line of the chiral order parameter $\Sigma $ (solid red) which can serve as a rough estimate of the BEC-BCS crossover inside the pion condensation phase, as explained below.

The RG-invariant pion condensate $\Delta $ in units of the scale parameter $M_0$ is represented as a contour plot with color shading in Fig.~\ref{fig:muI_Y_Delta}. A value of $\Delta/M_0 = 1$ (black) means that its value has reached that of the chiral order parameter in the vacuum, where $\Sigma = M_0$. At leading order, and approximately also in our RG-invariant MF calculation, 
$M^2 = \Sigma^2 +\Delta^2 $ stays constant
as the ground state rotates from $\Sigma = M_0$, $\Delta = 0 $ in the vacuum to $\Sigma \to 0 $, $\Delta \sim M_0 $
with increasing $x= 2\muI/m_\pi $ at vanishing temperature. The BEC-BCS crossover roughly occurs where $\Sigma \sim \Delta $. In the quark-meson model this is the point where the dynamically generated quark mass rotates from the original Dirac mass (when $\Sigma \gg \Delta$) to being predominantly of Majorana type (for $\Delta \gg \Sigma$). 

Moreover, the boundary of the pion condensation phase at finite temperature  in the RG-invariant MF calculation (solid green line) also compares remarkably well with that obtained on the lattice (light green band) in Refs.~\cite{Brandt:2017oyy,Brandt:2018omg}. 
Once more, the plot highlights how well the quark-meson model is able to describe the lattice results in our RG-invariant MF approach. The only region in parameter space where slight quantitative discrepancies can be seen is the corner of the pion-condensation phase in the transition region between the $\muI$-independent vertical phase boundary and the only weakly temperature dependent horizontal phase boundary. 

Although the remarkably rectangular shape of the pion-condensation phase boundary is thus not completely captured  by our quark-meson model description, it is worthwhile to note that the RG-invariant MF calculation here describes the lattice data considerably better than comparable previous model result in this regard.  Compared to the functional renormalization group (FRG) study in Ref.~\cite{Kamikado:2012bt}, for example, the RG-invariant mean-field calculation is conceptually closest to the extended mean-field (eMF) calculation where one integrates
the fermionic FRG flow for the effective potential at fixed meson field variables from a fixed UV cutoff $\Lambda$ down to the infrared. 
However, the eMF calculation of Ref.~\cite{Kamikado:2012bt} yielded a tricritical point around $\muI \sim 200$~MeV where the finite temperature phase boundary turned into a first-order transition, as had previously been  observed in an NJL-model calculation \cite{Andersen:2007qv} and originally predicted from  NLO $\chi$-PT for the analogous case in two-color QCD \cite{Splittorff:2002xn}. The most likely cause of this qualitative discrepancy in the eMF calculation is the finite UV cutoff $\Lambda $ whose effect for larger $T$ and $\muI $ might not be negligible but could in principle be eliminated in an RG-consistent extension along the lines of Refs.~\cite{Braun:2018svj,Gholami:2024diy} to remedy this. 

On the other hand, while there was no such tricritical point in the full FRG solution of Ref.~\cite{Kamikado:2012bt}, including the fluctuations from collective mesonic excitations to the effective potential at leading order in the derivative expansion, the resulting pion-condensation phase boundary was not nearly as rectangular as in Fig.~\ref{fig:muI_Y_Delta}.   
The same overall effects of fluctuations from the FRG were observed in the analogous phase diagram of two-color QCD \cite{Strodthoff:2011tz}, where the similarly horizontal phase boundary of diquark condensation \cite{Boz:2013rca} was later attributed to confinement effects as modeled by effective Polyakov-loop variables \cite{Strodthoff:2013cua}.  
While analogous FRG studies for QCD at finite isospin density will be important and interesting for the future, here we conclude that the main qualitative features of the corresponding phase diagram are very well captured by the simple quark-meson model in our RG-invariant MF calculation which is thus ready to serve as an effective description of the underlying dynamics and the competing order of chiral symmetry breaking versus pion condensation.

\begin{figure}
    \centering
    \includegraphics[width=\linewidth]{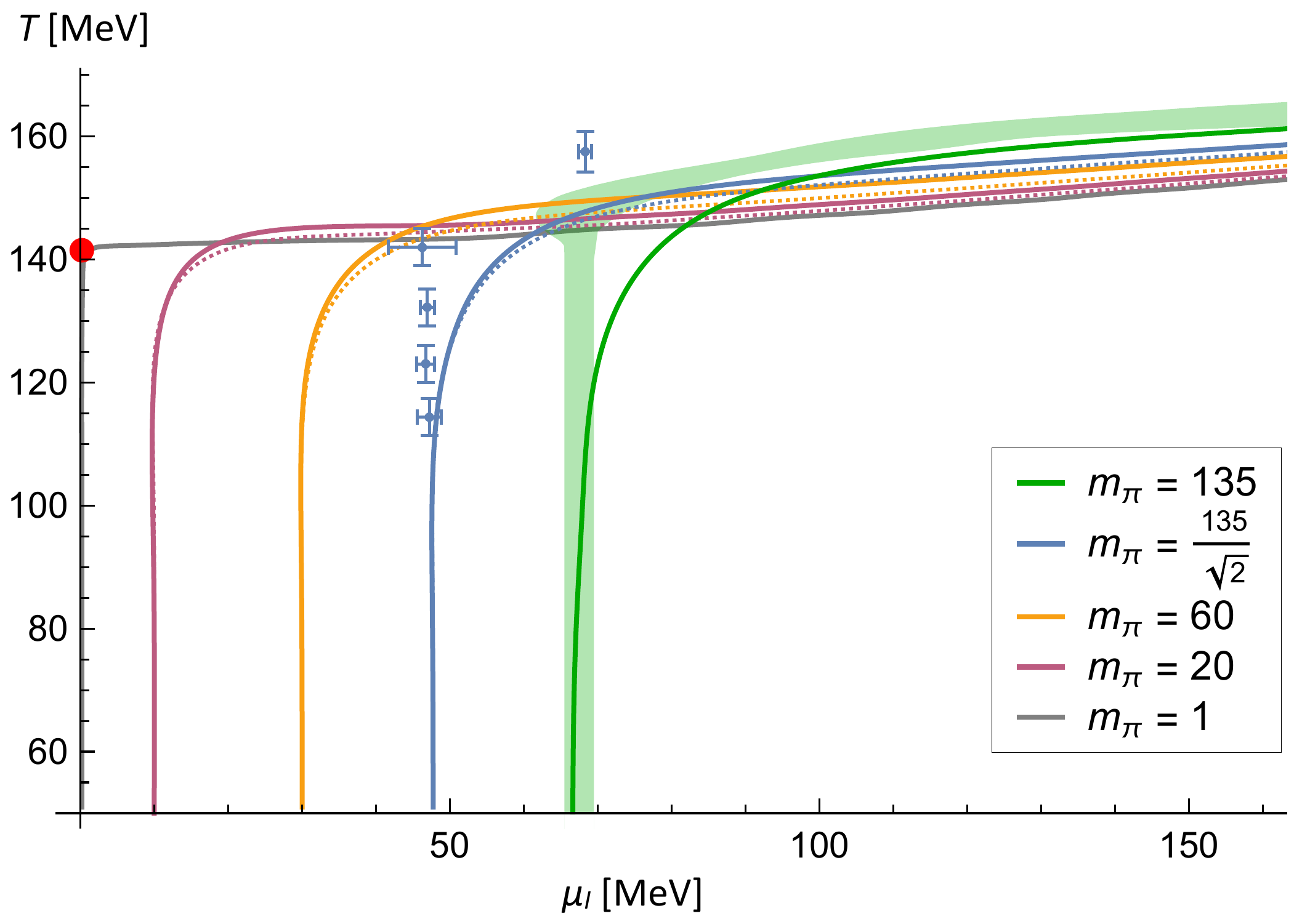}
    \caption{Pion condensation phase boundary towards the chiral limit, using our standard parameters $M_0=350$~MeV, $f_\pi=90$~MeV and $m_\sigma=470$~MeV. The red dot marks the two-flavor $O(4)$ chiral phase transition, here at $T_c=142$~MeV. Also shown are the lattice determination of the phase boundary of pion condensation at  $m_\pi=135$~MeV from Ref.~\cite{Brandt:2017oyy} (green band) and recent data for half the physical light-quark masses, corresponding to  $m_\pi=135/\sqrt 2$~MeV (blue points)~\cite{Brandt:2023kev,Brandt:2025hpy}.}
    \label{fig:muI_T_Delta}
\end{figure}

For the phase diagram at non-zero isospin chemical potential  this competing order becomes particularly interesting as one approaches the chiral limit, for early discussions within the NJL model, see Ref.~\cite{He:2005nk}, and the functional renormalization group, see Ref.~\cite{Svanes:2010we}. The problem is analogous to that in two-color QCD where diquark condensation at finite baryon chemical potential competes with chiral symmetry breaking in the chiral limit, as discussed in Refs.~\cite{Strodthoff:2011tz,vonSmekal:2012vx}.

Since pion condensation at $T=0$ sets in at $\muI=m_\pi/2$, in the chiral limit, where the pion mass vanishes, pion condensation occurs already for infinitesimal isospin chemical potentials $\muI>0$, spontaneously breaking the $O(2)$ subgroup of the isospin symmetry corresponding to  rotations in the charged pion plane. The limits $\muI \to 0 $ at $m_q = 0$ and $m_q\to 0 $ at $ \muI=0$ do not commute. More precisely, introducing a pion source $\hat\lambda$ as the external symmetry breaking field for  the $O(2)$ isospin rotations, this can be understood as follows: 

At $\muI =0$, the difference between $m_q\to 0$ at  $\hat\lambda = 0 $  and $\hat\lambda \to 0 $ at $m_q=0 $ is simply the alignment on the vacuum (coset) manifold of chiral symmetry breaking, with no effect on the universality class. 

At $m_q=0$, on the other hand,  any small but finite $\muI >0$ will lead to pion condensation in the limit $\hat\lambda \to 0$. 
Since changing the sign of $\hat\lambda $ changes that of the pion condensate $\Delta$ this leads to a first-order phase transition line at $\hat\lambda  = 0 $ along the temperature axis which ends in critical point of the three-dimensional $O(2)$ universality class at the corresponding Curie temperature $T_\Delta(\muI)$. In the limit $\muI \to 0 $, the latter may or may not coincide with the critical temperature $T_c$ of the two-flavor chiral phase transition.\footnote{$T_\Delta (0) $ can strictly speaking not be larger than $T_c$ because pion condensation brings chiral symmetry breaking along with it.}  
At $m_q=0$ there are thus essentially three conceivable scenarios depending on the behavior of   $T_\Delta (\muI)$ for $\muI\to 0$:

(i) If $T_\Delta (0) = 0 $, the pion condensate would have to vanish in this limit. This will certainly not happen, if we approach this limit with finite $\hat\lambda $ and send $\hat\lambda \to 0$ afterwards, which will produce chiral symmetry breaking in the twisted-mass direction of $\hat\lambda$, as explained above. 

(ii) If $0 < T_\Delta(0)  < T_c $, then there would be two distinct transitions when reducing the temperature from $T>T_c $ down to $T < T_\Delta $ (at $m_q=\muI=0$), the spontaneous breaking of chiral symmetry with the unbroken isospin subgroup at $T_c$, followed by the subsequent breaking of the isospin symmetry at some lower temperature $T_\Delta$. 

(iii) If $T_\Delta(0) = T_c $, on the other hand, then this marks a multicritical point. 

At least in our RG-invariant MF calculation we can answer this question. For $\muI =0 $ and $m_q\to 0$, the chiral phase transition of the two-flavor quark meson model is of $O(4)$ universality. For $m_q >0 $ on the other hand, pion condensation starts at $\muI = m_\pi/2$ and the transition is in the $O(2)$ universality class, as predicted for QCD~\cite{Son:2000xc}. In the limit $m_q\to 0 $ the two transitions meet in the point marked by the red dot in Fig.~\ref{fig:muI_T_Delta}. This corresponds to scenario (iii) where $T_\Delta(0) = T_c$, so that the red dot marks a multicritical point in the extended phase diagram including $m_q $ and $\muI $. 

To approach the chiral limit in our RG-invariant quark-meson model calculation, in principle, we simply need to reduce the pion mass in Eq.~\eqref{eq:mfOmegaRGphys} at fixed scale parameter $M_0$, eventually sending $m_\pi\to 0 $. From the formulae in the appendix, assuming that the bare parameters of the symmetric mesonic potential in Eq.~\eqref{eq:baresymmPot} remain unchanged, one can then in principle recompute the corresponding $\sigma$-mass parameter $m_\sigma$.  Here, we follow a slightly different strategy starting from the grand potential in Eq.~\eqref{eq:mfOmegaRGphys} and readjusting the $\sigma$-meson mass parameter to describe a dropping pseudo-critical temperature $T_{pc} $ of the chiral transition.  To illustrate the general trend, in Fig.~\ref{fig:muI_T_Delta} we do this in two different ways, assuming $O(4)$ scaling of $T_{pc} $ with the pion mass (solid lines) and mean-field scaling (dashed lines), with $T_{pc} \to T_c = 142$~MeV in either case.
\footnote{We have chosen the value of $T_c$ for the two-flavor quark-meson model slightly higher than the (2+1)-flavor lattice estimate of Ref.~\cite{HotQCD:2019xnw}, because the  pseudo-critical temperature $T_{pc} $, when defined via the peak of the susceptibility of the chiral condensate, in our model calculation at physical pion mass is also somewhat higher than current lattice estimates.}  

The boundary of the pion-condensation phase in Fig.~\ref{fig:muI_T_Delta}, for $m_\pi = 135$~MeV (green) is the same as in Fig.~\ref{fig:muI_Y_Delta} and includes the same (light green) band from the lattice determination. The result at half of the physical quark mass, corresponding to $m_\pi=135/\sqrt{2}$ (blue), is  compared to recent lattice data~\cite{Brandt:2023kev,Brandt:2025hpy}, shown as the blue data points in Fig.~\ref{fig:muI_T_Delta}. The plot shows that both for the model and the lattice data, the main effect of lowering the quark mass is to shift the phase boundary closer to the $\muI=0$ axis, maintaining its straight vertical section parallel to this axis. Further lowering the pion mass to $m_\pi= 60$~MeV (yellow), $m_\pi = 20$~MeV (purple) and, in particular, down to $m_\pi= 1$~MeV (gray) in the RG-invariant quark-meson  model calculation the pion-condensation phase boundary becomes more and more rectangular, also in the quark-meson model, and eventually runs parallel to the $\muI$-axis up to the sharp corner that builds up precisely at the point of the chiral phase transition (red dot), thus numerically confirming scenario (iii) with $T_\Delta(0) = T_c$, here.  
Moreover, the results in Fig.~\ref{fig:muI_T_Delta} can be combined into the exact same three-dimensional phase diagram in $m_q $, $\mu_I$ and $T$ as that sketched in Fig.~1 of Ref.~\cite{Strodthoff:2011tz} for two-color QCD at finite diquark-baryon density.
The qualitative agreement with the available lattice data provides reassuring evidence in favor of this scenario (iii) being realized in the chiral limit of full QCD as well.

\section{Summary and conclusions}

In this paper, we have presented a renormalization-group (RG) invariant 
formulation for the mean-field (MF) thermodynamics of the quark-meson model expressed entirely in terms of physical quantities such as $m_\pi$, $f_\pi $ and $m_\sigma$ together with one RG-invariant scale parameter $M_0 $ acting as the $\Lambda_\mathrm{QCD}$ of the model. 
This scale parameter is determined from the zero-temperature equation of state (EoS) in the low-energy region between the onset of pion condensation at isospin chemical potential $\muI = m_\pi/2 $  up to twice that, i.e.~up to $\muI \sim m_\pi$ matching lattice data. This zero-temperature EoS turns out to be completely insensitive to the $\sigma$-meson mass parameter of the effective model in this RG-invariant MF formulation, for all practical purposes. We can therefore independently adjust $m_\sigma $ to match the pseudo-critical temperature of the chiral transition at $\muI=0$ to the lattice data, which then completely fixes the model input. 

Using the matched parameters we have compared the results obtained from the model to the lattice data for the equation of state and the speed of sound at $T=0$ available in the literature~\cite{Brandt:2022hwy,Abbott:2024vhj}, together with NLO chiral perturbation theory~\cite{Andersen:2023ofv} and perturbation theory including contributions from the superconducting pairing gap~\cite{Fukushima:2024gmp}. We found good qualitative agreement with the lattice data for the speed of sound of Ref.~\cite{Brandt:2022hwy}. The model shows a tendency towards a peak at larger $\mu_I$ for the speed of sound which is in agreement with the results from Ref.~\cite{Abbott:2024vhj}, as shown in Fig.~\ref{fig:SoundSpeed_compare_3}. The plot shows that the RG-invariant quark-meson model with physical parameters is able to describe the presently available knowledge about the speed of sound in QCD surprisingly well over a wide range of chemical potentials, varying
over almost two orders of magnitude.

In the next step we compared the phase diagram in the plane of non-zero temperature and isospin chemical potential of our RG-invariant quark-meson model to the one obtained in lattice QCD in Ref.~\cite{Brandt:2017oyy}. For this comparison $m_\sigma$ has been tuned so that the model matches the pseudo-critical temperature at $\mu_I=0$. The resulting phase diagram is in remarkable agreement with the lattice QCD result, as shown in Fig.~\ref{fig:muI_Y_Delta}. This is also true for the almost rectangular boundary of the pion condensation phase, where comparable previous models often had problems to describe the findings of lattice QCD, e.g. Ref.~\cite{Kamikado:2012bt}.
In this context, future FRG studies for QCD at finite isospin density will be important and interesting.

Finally, we have investigated the behaviour of the pion condensation phase boundary in the approach to the chiral limit. Three possible scenarios have been discussed in the previous section. In our RG-invariant mean field calculation, the Curie temperature $T_\Delta$, the temperature where the pion condensate vanishes, becomes equivalent to the chiral transition temperature $T_c$ in the chiral limit for $\mu_I\to0$. The chiral phase transition point in the chiral limit at $T=T_c$ and $\mu_I=0$ therefore becomes a multicritical point. The behavior of the pion condensation phase boundary approaching the chiral limit is in qualitative agreement with recent lattice data~\cite{Brandt:2023kev,Brandt:2025hpy}, indicating that a similar scenario might be realized in the chiral limit of full QCD as well.

\begin{acknowledgments}
We thank Jens Oluf Andersen, Kenji Fukushima and Hosein Gholami for helpful discussions. 
This work was supported by the DFG (Collaborative Research Center CRC-TR 211 ``Strong-interaction matter under
extreme conditions'' - project no.~315477589), the Hungarian National Research, Development and Innovation Office (Research Grant Hungary 150241), the European Research Council (Consolidator Grant 101125637 CoStaMM), and the MKW NRW under the funding code NW21-024-A.
\end{acknowledgments}

\section*{Data availability}

The data that support the findings of this article are
openly available~\cite{data_pub}. The data from other publications for Figs.~\ref{fig:IsospinDensityData},~\ref{fig:IsospinandEnergyDensity} and~\ref{fig:SoundSpeed_compare} is available as ancillary files in the arXiv version of Ref.~\cite{Brandt:2022hwy}, for Fig.~\ref{fig:SoundSpeed_compare_3} the lattice data is available as described in Ref.~\cite{Abbott:2024vhj} and for Fig.~\ref{fig:muI_T_Delta} it is available in the ancillary files accompanying the arXiv version of Ref.~\cite{Brandt:2025hpy}.

\appendix
\section{Regularization and Renormalization}
The mean-field renormalization for vanishing isospin density is a standard procedure \cite{Skokov:2010sf}, it requires renormalizing only the quartic meson coupling $\lambda$ to absorb the logarithmic divergence proportional to $M^4$. With pion condensation, on the other hand, an additional logarithmic divergence $\propto \Delta^2 \muI^2$ arises \cite{Andersen:2018osr,Folkestad:2018psc,Ayala:2023cnt}. This independent additional UV divergence structure requires meson field renormalization.
To regularize the corresponding terms given in Eq.~\eqref{eq:TaylorVacmu}, it is convenient to use prescriptions analogous to those for $\zeta$-function regularization in studies of Casimir forces from vacuum fluctuations, based on integrals of the form
\begin{equation}
\int p^2 dp  \, \big( p^2 + M^2)^{(1-s)/2} = \frac{M^{4-s} \sqrt{\pi}\,\Gamma\big(\frac{s-4}{2}\big)}{4 \Gamma\big(\frac{s-1}{2}\big) } \,. \label{eq:A1}
\end{equation}
With these we obtain
\begin{align}
    \int p^2 dp \,  \sqrt{p^2 + M^2} &= - \frac{M^4}{8 \epsilon} \, + \label{eq:A2} \\ 
    & \hspace{-.8cm}\frac{M^4}{32} \Big( 1 -4 \ln 2 + 4 \ln\big(M/\muRG\big) \Big) + \mathcal O (\epsilon) \, ,\nonumber \\
 \int p^2 dp \,  \frac{1}{\sqrt{p^2 + M^2}^3}  &= \frac{1}{\epsilon}  + \ln 2 - 1 -\ln\big(M/\muRG\big) + \mathcal O (\epsilon) \, ,\nonumber
\end{align}
where we have used $s = \epsilon\to 0^+$ in the first line and $s-4 = \epsilon \to 0^+ $ in the second, and $\muRG $ is the renormalization scale. Thus, to absorb the UV divergences in $\Omega^0_q(\muI^2)$, we use $Z_u  = Z_g^2/Z_\lambda $ and $Z_v = Z_g$ and identify
\begin{align}
   Z_u  &=  1 + \frac{N_c g^4}{\pi^2 \lambda} \Big(\ln(\Lambda/\muRG)  + \ln 2 - \frac{1}{4}\Big)\, ,\label{eq:Zu} \\
 Z_v &=  1 + \frac{N_c g^2}{2\pi^2} \Big(\ln(\Lambda/\muRG)  + \ln 2 - 1 + C_\phi \Big)\, , \label{eq:Zv}
\end{align}
where we have replaced $1/\epsilon \to \ln(\Lambda/\muRG) $  (which corresponds to their form when using a three-dimensional momentum cutoff $\Lambda $ instead of \eqref{eq:A1} in Eqs.~\eqref{eq:A2}), in order to emphasize the renormalization-scale $\muRG $ dependence of these factors, and to readily compute the corresponding $\beta$-functions in \eqref{eq:betau} and \eqref{eq:betav}.
With this definition of the renormalization constants, our effective renormalized potential, expressed in terms of the couplings $u$ and $v$, reads
\begin{align}
    \Omega(T,\muq,\muI) &= \frac{m^2}{2} (\rho^2+ d^2) +\frac{1}{4u}
    g^4 (\rho^2+d^2)^2 - c \sigma \nonumber\\
    & - 2\muI^2  \frac{g^2 d^2}{v}   + R^\mathrm{vac}_q(\muI^2) + \Omega^R_q(T,\muq,\muI) \nonumber\\
     &- \frac{N_c}{8\pi^2} \, g^4(d^2+\rho^2)^2 \ln\Big(\frac{g^2(d^2+\rho^2)}{\muRG^2}\Big)  \nonumber\\
     &+ \frac{N_c}{2\pi^2} \, \muI^2 g^2d^2 \left( \ln\Big(\frac{g^2(d^2+\rho^2)}{\muRG^2}\Big) + 2 C_\phi\right) \, \label{eq:mfOmegarenf},
\end{align}
where the UV finite quark contributions can explicitly be written as
\begin{widetext}
\begin{align}
     \Omega_q(T,\muq,\muI) &\equiv
    \Omega_q(T,\muq,\muI) -   \Omega_q^{0}(0) - \muI^2 \frac{\partial }{\partial \muI^2 } \Omega_q^{0}(\muI^2)\big\vert_{\muI^2=0} \label{eq:mfOmegaR} \\
    &= - 2 T N_c \idp \sum_\pm \bigg( \ln\Big( 1+ e^{-\frac{|E_\pm -\muq| }{T}} \Big) +\ln\Big( 1+ e^{-\frac{|E_\pm +\muq| }{T}} \Big) \bigg) \nonumber \\ 
    &\hskip .4cm + 2 N_c \idp \sum_\pm \theta\big(|\muq| - E_\pm \big) \big(E_\pm - |\muq|\big) + \int_0^{\muI^2} \!
    dx\, \bigg(   \nonumber \frac{d }{dx } \Omega_q^{0}(x) + \frac{N_c}{\pi^2} \int p^2 dp \, \frac{g^2 d^2 }{\sqrt{p^2 + g^2 (d^2+\rho^2)}^3}  \bigg)\, .
\end{align}
\end{widetext}
It is then manifestly renormalization group invariant, satisfying the renormalization group equation
\begin{align}
    \Big(\muRG\frac{\partial} {\partial\muRG } + \beta_u(u) \frac{\partial} {\partial u }+ \beta_v(v) \frac{\partial} {\partial v } \Big) \, \Omega(T,\muq,\muI) &= 0\, ,
\end{align}
with the $\beta$-functions from Eqs.~\eqref{eq:betau} and \eqref{eq:betav}.
An explicitly renormalization-group (RG) invariant form of the effective potential is obtained by introducing RG invariant variables $\tilde c \equiv c/g $, $\tilde\sigma = g\sigma $ and $\tilde m^2= m^2/g^2 $. For the field invariants, we again use $M$ and $\Delta$. Then, the effective potential still depends on $u$ and $v$, albeit in the RG invariant combination $2/v - 1/u$. We can eliminate this dependence, noting that 
\begin{align}
    \ln\big(\muRG/\muRG_0) &= \int_{u_0}^u \frac{dl}{\beta_u(l)} = - \frac{\pi^2}{N_c} \Big(\frac{1}{u} - \frac{1}{u_0} \Big)\\ 
    &= \int_{v_0}^v \frac{dl}{\beta_v(l)} = - \frac{2\pi^2}{N_c} \Big(\frac{1}{v} - \frac{1}{v_0} \Big) \, ,\nonumber
\end{align}
where $u_0 = u(\muRG_0)$ and $v_0 = v(\muRG_0)$ are the values of the couplings $u=g^4/\lambda$ and $v=g^2$ at the reference scale $\muRG_0 = M_0$. Therefore 
\begin{equation}
    \frac{2}{v} - \frac{1}{u} + \frac{1}{u_0} = \frac{2}{v_0} \, .
\end{equation}
$u_0$ is determined from the requirement that $\Omega^R(0,0,0) $ is minimized  for $M=M_0 $, leading to
\begin{align}
    u_0 = \big(\tilde c/M_0^3  - \tilde m^2/M_0^2 + N_c/(4\pi^2) \big)^{-1} \, ,
\end{align}
where $M_0 = \sqrt{v_0} \, \sigma_0$. With $\sigma_0 = f_\pi$ in our mean-field approximation, this in turn then also fixes 
\begin{align}
    v_0 = M_0^2/f_\pi^2 \,.
\end{align}
For the pion mass, we simply obtain
\begin{align}
    m_\pi^2 = v_0 \, \tilde c/M_0 
    = \tilde c M_0/f_\pi^2  = c/\sigma_0 \, ,
\end{align}
which corresponds to the Gell-Mann-Oakes-Renner relation of the model,
\begin{equation}
\tilde c M_0 = c \sigma_0 = m_\pi^2 f_\pi^2  \, .
\end{equation}
Finally, the $\sigma$ meson mass is given by 
\begin{align}
    m_\sigma^2 &= v_0 \big( -2 \tilde m^2 + 3\tilde c/M_0 - N_c M_0^2/\pi^2\big) \, .
\end{align}
Using these relations to eliminate $u_0$ and $v_0 $  as well, the effective potential in its final form writes
\begin{widetext}
\begin{align}
    \Omega^R(T,\muq,\muI) &=  \frac{1}{2} \, \Big(\frac{(3m_\pi^2 - m_\sigma^2)f_\pi^2}{2M_0^2}  -\frac{N_c}{2\pi^2} \,M_0^2 \Big)    M^2  +\frac{1}{4} \,
     \Big(
     \frac{(m_\sigma^2- m_\pi^2)f_\pi^2}{2M_0^4}
     + \frac{3N_c}{4\pi^2}  \Big) M^4  - \frac{m_\pi^2f_\pi^2}{M_0} \, \sqrt{M^2-\Delta^2} 
    \nonumber\\[4pt]
     &   \hskip 1cm 
     - \frac{N_c}{4\pi^2} \, M^4  \ln\big(M/M_0\big)  + \frac{N_c}{\pi^2} \, \muI^2 \Delta^2 \Big(\ln\big(M/M_0\big) +C_\phi\Big)  + R^\mathrm{vac}_q(M^2,\Delta^2;\muI^2) \\[6pt]
     & \hskip -1cm - 2 T N_c \idp \sum_\pm \bigg( \ln\Big( 1+ e^{-\frac{|E_\pm -\muq| }{T}} \Big) +\ln\Big( 1+ e^{-\frac{|E_\pm +\muq| }{T}} \Big) \bigg) 
 + 2 N_c \idp \sum_\pm \theta\big(|\muq| - E_\pm \big) \big(E_\pm - |\muq|\big) \, .
    \nonumber 
\end{align}
\end{widetext}
With $C_\phi = - F^0_q\big(m_\pi/(2M_0)\big)$, as explained in the main text, determined from
\begin{align}
    F^0_q(x) &=  \frac{2 \sqrt{1-x^{2}}}{x} \bigg( \arctan\bigg(\frac{x}{\sqrt{1-x^{2}}} \bigg) \\
    &\hskip 1.4cm+ \arctan\bigg(\frac{1-x}{ \sqrt{1-x^{2}}} \bigg)  
- \frac{\pi}{4} \bigg) - 1 \, .\nonumber
\end{align}
As discussed above, the RG invariant $M_0$ defines the overall scale. If we state all dimensionful quantities in units of $M_0$, then there is in principle no free parameter other than this overall scale. In practice, of course, unlike the physical pion mass $m_\pi$, the mass parameter $m_\sigma $ of the very broad $\sigma$-meson resonance is only rather poorly constrained. In the quark-meson model it is well known to control the temperature of the chiral transition at vanishing chemical potentials, and hence effectively acts as a second parameter. In our case this leads to $m_\sigma \approx 470$~MeV which, encouragingly, is of the expected order for the broad two-pion resonance pole at $m_\sigma\simeq ( 450 - \iu\, 275 )$~MeV  \cite{Pelaez:2015qba}.  

\section{Relation to \boldmath $\overline{MS}$ Scheme}

It is far from obvious how to compare our RG-invariant MF potential in Eq.~\eqref{eq:mfOmegaRGphys} in detail to that derived in Refs.~\cite{Andersen:2018osr,Adhikari:2018cea}, cf.~Eq.~(26) in \cite{Adhikari:2018cea}.
The most important difference is the renormalization scheme. On the one hand, we use an on-shell scheme where the scale parameter is defined as the (RG-invariant) fermion mass, $M_0 = g\sigma $, and the meson field renormalization is fixed such that our mass parameters agree with the vacuum curvature masses as identified from the (RG-dependent) mesonic two-point functions $\Gamma^{(2)}(p)$ for vanishing momentum in the vacuum (with $\mu_I=0$) at the reference scale, i.e.
\begin{equation}
   \Gamma^{(2)}_\pi(0) = \frac{g^2}{g_0^2} \, m_\pi^2\, , \;\;  \Gamma^{(2)}_\sigma(0) = \frac{g^2}{g_0^2} \, m_\sigma^2\, , \label{eq:curvemass}
\end{equation} 
see, e.g., Eq.~\eqref{eq:rencond_pion}. Refs.~\cite{Andersen:2018osr,Adhikari:2018cea}, on the other hand, use a modified minimal subtraction ($\overline{MS}$) scheme based on dimensional regularization in $d=3-2\epsilon $ spatial dimensions. To verify that the logarithmic divergences are identical, it is therefore necessary to replace $2\epsilon $ in Refs.~\cite{Andersen:2018osr,Adhikari:2018cea} by $\epsilon\,  \widehat=  \ln(\Lambda/\muRG)  $, here. With this identification, the logarithmic divergences in the renormalization constants of the equations in (27) of Ref.~\cite{Andersen:2018osr} are readily verified to agree with the two independent ones in Eqs.~\eqref{eq:Zu} and \eqref{eq:Zv} above. The finite subtractions differ, however. To relate the RG-invariant scale parameter $M_0 = g\sigma $ used here to the analogous $\Lambda_0 $
of the $\overline{MS}$ scheme in Refs.~\cite{Andersen:2018osr,Adhikari:2018cea}, we can use Eq.~(B18) of the same Appendix B in either of the two references to conclude
\begin{equation}
    \ln\Big(\frac{M_0^2}{\Lambda_0^2}\Big)  = F(m_\pi^2) + m_\pi^2 F'(m_\pi^2)\, , 
\end{equation}
where the function $F(x) $ from (B8) of App.~B in \cite{Andersen:2018osr,Adhikari:2018cea} (noting a typo in their (B9), where $r^2$ in the denominator should probably be $p^2$) is given by,
\begin{equation}
    F(m^2) = 2 - 2\sqrt{\frac{4 M_0^2}{m^2}-1 }  \, \arctan\Big(1/\sqrt{\frac{4 M_0^2}{m^2}-1 }\Big)\, . \nonumber
\end{equation} 

\newpage
\noindent 
The meson wavefunction renormalization here, with
\begin{equation}
    Z_\phi = Z^{-1}_v =  1- \frac{N_c g^2}{2\pi^2} \Big(\frac{1}{\epsilon}  + \ln 2 - 1 + C_\phi \Big)\, , \label{eq:Zphi}
\end{equation}
also differs from the $\overline{MS}$ scheme of \cite{Andersen:2018osr,Adhikari:2018cea} by finite terms. While with our second renormalization condition the meson mass parameters $m_\pi^2$ and $m_\sigma^2$ agree with the curvature of the effective vacuum potential at the reference scale, cf.~\eqref{eq:curvemass}, we can identify, e.g., most easily from Eq.~(C9) in \cite{Andersen:2018osr} that the corresponding pion curvature mass parameter of Refs.~\cite{Andersen:2018osr,Adhikari:2018cea} is given by
\begin{equation}
   \Gamma^{(2)}_\pi(0) = \frac{g^2}{g_0^2} \, m_\pi^2\, 
  \Big( 1 - \frac{g_0^2 N_c}{4\pi^2} m_\pi^2 F'(m_\pi^2)\Big) \, . \label{eq:curvemassAK}
\end{equation} 
where we have inserted $m_q^2 = g^2_0f_\pi^2 $ in Eq.~(C9) of  \cite{Andersen:2018osr}.
This implies that our pion curvature mass parameter at the reference scale needs to be replaced by
\[ m_\pi^2 \to m_\pi^2  \Big( 1 - \frac{g^2_0 N_c}{4\pi^2} m_\pi^2 F'(m_\pi^2)\Big) \]
when comparing our scheme with that of Refs.~\cite{Andersen:2018osr,Adhikari:2018cea}.
Numerically, the finite renormalization by which the two differ, with our typical choice of parameters, throws in a factor of $\approx 0.97$ in the brackets here, and hence only about $1.5\%$ difference in the corresponding pion masses. A relation between the respective curvature mass parameters $m_\sigma^2 $ for the $\sigma$-meson field can be derived analogously. 

Note however that both the pion decay constant $f_\pi$ as well as the fermion mass $m_q = g \sigma$ are RG-invariant while $g\equiv g(\nu)$ necessarily depends on the renormalization scale $\nu$, if meson field renormalization (inverse to that of the Yukawa coupling)  is required as in the renormalized mean-field potential of the quark-meson model with finite isospin chemical potential. This is why we only ever identified $\sigma_0 = f_\pi$ and hence $m_q = g_0 f_\pi $ at the reference scale $\nu_0 $, while $m_q = g f_\pi  $ appears to be assumed in Refs.~\cite{Andersen:2018osr,Adhikari:2018cea} which needs to be amended in presence of meson field renormalization. An exact equivalence between our RG-invariant mean-field potential in Eq.~\eqref{eq:mfOmegaRGphys} and that of Ref.~\cite{Adhikari:2018cea} is therefore difficult to establish.

\newpage

\end{document}